\documentclass[12pt,preprint]{aastex}

\makeatother

\lefthead{Rigopoulou et al..~}

\slugcomment{to appear in ApJ}
\shortauthors{Rigopoulou et al.}

\shorttitle{FIR Spectroscopy of Intermediate Redshift ULIRGs}
\shortauthors{Rigopoulou et al.}

\begin{document}

\title{{\it Herschel} Observations of Far-Infrared Cooling Lines in intermediate Redshift (Ultra)-luminous Infrared Galaxies\thanks{Herschel is an ESA space observatory with science instruments provided by European-led Principal Investigator consortia and with important participation from NASA}}

\author{D. Rigopoulou\altaffilmark{1,2},
R. Hopwood\altaffilmark{3},
  G.E. Magdis\altaffilmark{1}, 
  N. Thatte\altaffilmark{1},
  B.M. Swinyard\altaffilmark{2,4} 
 D. Farrah\altaffilmark{5}, 
J-S. Huang\altaffilmark{6},
A. Alonso-Herrero\altaffilmark{7}, 
J.J. Bock\altaffilmark{8,9}, 
D. Clements\altaffilmark{3},
A. Cooray\altaffilmark{10,9},
M.J. Griffin\altaffilmark{11},
S. Oliver\altaffilmark{12},
C. Pearson\altaffilmark{2,13},
D. Riechers\altaffilmark{14},
D. Scott\altaffilmark{15}
A. Smith\altaffilmark{12},
M. Vaccari\altaffilmark{16},
I. Valtchanov\altaffilmark{17},
L. Wang\altaffilmark{12}}






\altaffiltext{1}{Department of Physics, University of Oxford, Keble Road, Oxford, OX1 3RH, UK}
\altaffiltext{2}{RAL Space, Science \& Technology Facilities Council, Rutherford Appleton Laboratory, Didcot, OX11 0QX, UK}
\altaffiltext{3}{Physics Department, Imperial College London, South Kensington
Campus, London SW7 2AZ, UK }
\altaffiltext{4}{Dept. of Physics \& Astronomy, University College London, Gower
St, London, WC1E 6BT, UK  }
\altaffiltext{5}{Department of Physics, Virginia Tech, Blacksburg, VA 24061, USA}
\altaffiltext{6}{Harvard-Smithsonian Center for Astrophysics, 60
  Garden Str.,  Cambridge, MA02138, USA}
\altaffiltext{7}{Instituto de Fisica de Cantabria, CSIC-UC, 39006
  Santander, Spain}
\altaffiltext{8}{California Institute of Technology, 1200 E. California Blvd., Pasadena, CA 91125, USA}
\altaffiltext{9}{Jet Propulsion Laboratory, 4800 Oak Grove Drive,  Pasadena, CA 91109, USA}
\altaffiltext{10}{Department of Physics \& Astronomy, University of California, Irvine, CA 92697, USA}
\altaffiltext{11}{School of Physics and Astronomy, Cardiff University,
  Queens Buildings, The Parade, Cardiff CF24 3AA, UK }
\altaffiltext{12}{Astronomy Centre, Department of Physics \& Astronomy, University of Sussex, Brighton BN1 9QH, UK} 
\altaffiltext{13}{Department of Physical Sciences, The Open University, Milton Keynes MK7 6AA, UK }
\altaffiltext{14}{Department of Astronomy, Cornell University, 220 Space Sciences Building, Ithaca, NY 14853, USA}
\altaffiltext{15}{Department of Physics and Astronomy, University of
  British Columbia, 6224 Agricultural Road, Vancouver, B.C. V6T1Z1, Canada}
\altaffiltext{16}{Astrophysics Group, Physics Department, University
  of the Western Cape, Private Bag X17, 7535 Bellville, Cape Town,
  South Africa}
\altaffiltext{17}{Herschel Science Centre, European Space Astronomy
  Centre, Villanueva de la Canada, E-28691 Madrid, Spain}

\begin{abstract}
We report the first results from a spectroscopic survey of the 
[CII] 158$\mu$m line from
a sample of intermediate redshift (0.2$<z<$0.8) (ultra)-luminous
infrared galaxies, (U)LIRGs (L$_{\rm IR}>$10$^{11.5}$L$_{\rm \odot}$), using the SPIRE-Fourier 
Transform Spectrometer
(FTS) on board the {\it Herschel}  Space Observatory. This is the first
survey of [CII] emission, an important tracer of star-formation,  at a
redshift range where the star-formation rate density of the Universe increases rapidly.
We detect strong [CII] 158$\mu$m line emission from over 
80\% of the sample. We find that the [CII] line is luminous,
in the range (0.8$-$4)$\times$10$^{-3}$ of the far-infrared continuum
luminosity of our sources, and appears to arise from
photodissociation regions on the surface of molecular clouds. The
$L_{\rm [CII]}/L_{\rm IR}$ ratio in our intermediate redshift
(U)LIRGs is on average $\sim$10 times larger than that of local ULIRGs.
Furthermore, we find that the $L_{\rm [CII]}/L_{\rm IR}$  and  $L_{\rm
 [CII]}/L_{\rm CO(1-0)}$ ratios in our sample 
are similar to those of local normal  galaxies and  high-$z$
star-forming galaxies. 
ULIRGs at $z\sim 0.5$ show many similarities to the properties of
  local normal and high-z star forming galaxies.
Our findings strongly suggest
that rapid evolution in the properties of the star forming regions of luminous infrared
galaxies is likely to  have occurred in the last 5 billion years.

\end{abstract}

\keywords{infrared: galaxies --- infrared: ISM --- galaxies: starburst}


\section{Introduction}

Luminous (10$^{11}<L_{\rm IR(8-1000)}$$<$10$^{12}$ L$_{\rm \odot}$) and ultra-luminous IR galaxies 
($L_{\rm IR(8-1000)}>$10$^{12}$ L$_{\rm \odot}$, (U)LIRGs) are amongst the most important
populations in studies of galaxy evolution. The origin of their
extreme luminosities has been the focus of debate since their
discovery by IRAS
however, it is now widely accepted that local (z$<$0.26)  ULIRGs are primarily
powered by star-formation (e.g.~Genzel et al.\,1998, Rigopoulou et
al.\,1999, Farrah et al.\,2007).  

While ULIRGs in the local Universe are rare (e.g. Lagache et al.\,2005), they contribute
significantly to
the total IR energy density from redshifts 0.5 and above (e.g.~Le
Floc'h et al.\,2005, Rodighiero et al.\,2010). At redshifts $z>$2 submillimetre galaxies (SMGs)
are considered to be the more luminous counterparts of local ULIRGs
(e.g.~Blain et al.\,2002). There are many indications however, that
local ULIRGs differ systematically from their high redshift
counterparts. 
Many authors
 (e.g.~Papovich et al.\,2007, Farrah et
al.\,2008, Muzzin et al.\,2010, Swinbank et al.\,2010) have found that
high-z ULIRGs have different Spectral Energy Distributions (SEDs), mid-infrared properties
and extent of star-forming regions when compared to local ones.
More recently,
Rujopakarn et al.\,(2011) compared physical scales of the star-forming
regions and concluded that high-$z$ ULIRGs
are more akin to local star-forming galaxies rather than local ULIRGs.

The fine structure line 
[CII] at 158$\,\mu$m is one of  the brightest emission lines in the
spectra of galaxies. [CII] traces gas exposed to 
far-ultraviolet (FUV) photons from OB stars with energies greater than 11.3 eV, the ionisation potential
of C$^{0}$. In these photodissociation regions (PDRs), atomic,
molecular, hydrogen and electrons can
collisionally excite the ground state of C$^{+}$ ions producing [CII]
which cools the gas. 
Early [CII] detections with the Kuiper Airborne
Observatory (KAO) in nearby galaxies showed that the
line was bright, 0.1$-$1\% of the observed  far-infrared (FIR)  luminosity (e.g. Stacey et
al.\,1991, Crawford et al.\,1985). Subsequent observations with the
{\it Infrared
Space Observatory (ISO)} confirmed these findings but highlighted
a deficit of the [CII] line in the highest luminosity systems such as
local ULIRGs (e.g.~Luhman et al.\, 1998, 2003).

Since most local ULIRGs are thought to be star-formation
dominated (e.g.~Genzel  et al.\, 1998, Rigopoulou et al.\, 1999) the [CII]-deficit appeared at odds
with previous results and a number of explanations were put forward, including size of
[CII] emitting regions, metallicity and dust content
(e.g. Luhman et al.\,1998, 2003). But recent detections of [CII] in
luminous z$>$1.5 star-forming systems with {\it Herschel} (e.g. Ivison et
al. 2010, Valtchanov et al.\,2011, George et al.\,2013) and other
ground-based facilities (e.g. Hailey-Dunsheath et al.\,2010, Stacey et
al.\, 2010) revealed that their [CII]/FIR luminosity ratios are similar to
local star-forming galaxies, much above the median values found for
local ULIRGs. 
More recently,
Gracia-Carpio et al.~\,(2011) found that the
[CII]$/$L$_{\rm FIR}$ ratio is inversely
proportional to L$_{\rm FIR} /$ M$(H_2$) for a fixed L$_{\rm FIR}$ for
a sample of local starbursts and (U)LIRGs.

 Here we present the first results of a
survey of [CII] in a sample of 0.2$<$$z$$<$ 0.8
(U)LIRGs. 
The redshift range 0.2$<$$z$$<$ 0.8
represents a crucial phase in galaxy evolution: it is exactly in this range that the star formation density
of the Universe increases steeply, becoming essentially flat at
$z$$>$1.5 (e.g. Magnelli et al.\, 2013, Bouwens et al.\,2009).  [CII] observations of galaxies
in this cosmic epoch will establish the long-sought link between the local and
high-$z$ Universe and allow us to form a benchmark for future studies of [CII] at higher redshifts.

\section{Sample Selection \& Observations}

\subsection{The Sample}

The primary goal of the survey is to investigate the
properties of the ISM and in particular
whether intermediate redshift (U)LIRGs are [CII]-deficient like their
local counterparts. To construct the present sample we employed the
{\it Herschel} Multi-tiered Extragalactic Survey (HerMES, Oliver et al.\,2012)
photometric catalogues 
produced using a
prior source extraction 
based on the position of known 24$\,\mu$m
sources (XID, Roseboom et al.\,2012). We
searched for sources that
satisfied the following two criteria: (1) $S_{\rm 250} >$ 150\,mJy, a limit
imposed to ensure detection of the source against the background
emission of the telescope at 80 K and, (2) redshift in the range 0.2$<$z$<$0.8 so that we could
observe at least one of the primary cooling lines [CII]
158$\,\mu$m, [NII] 205 $\,\mu$m and [CI] 307$\,\mu$m in the
194$-$671$\,\mu$m range.
These criteria resulted in the selection
of 22 (U)LIRGs with $L_{\rm IR(8-1000)}$$>$10$^{11.2}$L$_{\rm \odot}$, 17 of them with
confirmed spectroscopic redshifts ($z_{\rm spec}$). 

We have supplemented the far-infrared spectroscopic data with
single-dish CO data using the IRAM 30 m and ESO APEX telescopes and,
spatially resolved optical Integral field spectroscopic data using the
Oxford-SWIFT IFU (Thatte et al.\,2010) on Palomar. The present Letter
focuses on 12 intermediate redshift (U)LIRGs with $z_{\rm spec}$ for which the [CII]
158$\,\mu$m line falls in the SPIRE-FTS range. We note that the [OI]63 $\mu$m line does not fall in the FTS
range,  hence we assume that [CII] is the primary cooling line in
our sample. Measurements of the full
sample and reports of
the ancillary datasets and measurements will be presented in Magdis et al. (2014).

\subsection{SPIRE-FTS Spectroscopy}

The (U)LIRGs in our sample were observed with the Spectral and
Photometric Imaging REceiver 
(SPIRE; Griffin et al.\,2010) Fourier Transform Spectrometer (FTS)
on board the {\it Herschel} Space Observatory (Pilbratt et al.\,2010), between March
2012 and January 2013. The FTS observed 100 repetitions (13320 seconds total integration
time) on each target in single pointing,
high spectral resolution (0.048\,cm$^{-1}$) mode, with sparse spatial 
sampling. 
The SPIRE-FTS measures the Fourier transform 
of the spectrum of a source using two bolometer detector arrays, simultaneously 
covering wavelength bands of 194$-$313$\,\mu$m (SSW) and 303$-$671$\,\mu$m (SLW). 
 All (U)LIRGs in the sample are point-like, given the beam size in SSW and the distances.
A typical spectrum around the [CII] line is shown in Fig.~1. 
All ULIRGs in our sample are extremely faint targets for the FTS
and, require post-pipeline processing beyond the standard
reduction. A detailed account of the procedure used can be found in Hopwood et al.\,(2013). In brief, we
used the standard FTS pipeline (Fulton et al. 2013b in prep) in HIPE (Ott et
al.\,2010) version 11 to reduce the data. Within the pipeline steps, a
bespoke Relative Spectral Response Function (RSRF) 
constructed from selected long dark sky observations 
was applied if it improved the noise in the point source
calibrated product (level-2 product).

Two methods can be used to remove residual background 
in the level-2 spectra. Firstly, spectra from detectors around the central
``on-source'' detector were selected and their average spectra
subtracted from the central detector.
The second method involves subtracting a dark sky 
spectrum observed on the same operational day. 
We found the off-axis subtraction provided the best reduction in residual for all 
but one observation.
Once the optimum reduction method was determined and the 
level-2 spectra obtained, these were examined for spectral features. 
Since random noise in an FTS spectrum can easily mimic a faint
line  we used the jackknife technique as a reliability test to minimise spurious detections.
In brief, each unaveraged level-2 spectrum, of 200 scans, was split into sequential 
subsets. This was repeated for subsets of decreasing number of scans.
By plotting the averaged subsets at the expected line positions and making a visual comparison 
over all subsets, an assessment of the presence of an expected line
was made.

Once a line was assessed as real, a bootstrap method was used to measure the line 
flux. For any given observation, scans were randomly sampled until the number in the 
parent population (of 200) was reached. These random scans were then averaged and 
the line measurements made by fitting a sinc function, 
or, if the line was partially resolved  (in six cases) we employed a sinc convolved with a 
Gaussian. We repeated the process 10,000 
times for each observation and fitted a Gaussian to the resulting line flux distribution to 
obtain the mean line flux. 
The standard deviation was taken as the associated 1$\sigma$ 
uncertainties. Random frequency positions were also selected and the same process repeated to 
give comparison distributions and provide a second reliability check. Due to the nature 
of FTS random and systematic noise, the distributions obtained for 
random frequency positions are indistinguishable to those for faint lines ($<2\,\sigma$), however, a 
background level can be established with the bootstrapped results for these randomly 
selected positions, above which a real spectral feature is strongly suggested. For any
line found to be below $2\,\sigma$, the bootstrapped flux 
density is taken as an upper limit, but only 
if the presence of a [CII] line is supported by the jackknife and other visual checks.

\section{ Results  }

\subsection{The $L_{\rm [CII]}/L_{\rm IR}$ Relationship}

We have detected  [CII] line emission ($>$3$\sigma$) from 10 of the 12
ULIRGs with $z_{\rm spec}$.  The [CII] line is bright in all detected
sources, in the range (0.6-2.6)$\times$10$^{9}$L$_{\rm \odot}$.
The brightest source, with a [CII] line
luminosity of 5.7$\times$10$^{10}$ L$_{\rm \odot}$ has been identified
with a $z$$=$2.31 luminous galaxy merger (XMM01, Fu et al. 2013)
although the lens galaxy was originally selected.
The two sources with the
lowest [CII] line luminosity both contain spectroscopically confirmed
AGN (Houck et al. 2003) but are not classified as QSOs.

The parameter {\it R} $=L_{\rm [CII]}/L_{\rm FIR}$, defined as the ratio of the [CII] line luminosity
to the far-infrared continuum luminosity $L_{\rm FIR}$\footnote
{defined as the luminosity between 42.5$-$122.5$\,\mu$m} can be used to probe the strength of
the ambient radiation field ($G_{0}$, e.g. Kaufman
et al. 1999) under the PDR paradigm. 
Since the emergent FIR intensity is directly
proportional to the underlying radiation field and,
L$_{\rm[CII]}$ is only weakly dependent on {\it G$_{0}$} it follows that
their ratio {\it R} is inversely proportional to {\it G$_{0}$}. Low values
of {\it R}
would imply a hard underlying radiation field.
Despite local ULIRGs being predominantly starburst dominated (e.g. Rigopoulou et
al. 1999, Farrah et al. 2007) their ratio
{\it R}=$L_{\rm [CII]}/L_{\rm FIR}$  is surprisingly low, less than 1\%
(e.g. Luhman et al. 2003, Farrah et al. 2013). In contrast, high
redshift ULIRGs at z$\sim$1$-$2 (e.g. Stacey et al. 2010) and z $>$ 2
lenses discovered by {\it Herschel} (e.g. Ivison et al 2010, Valtchanov et
al. 2011) have R ratios similar to those of local star-forming
galaxies. 

In Fig.~2 we plot the observed {\it R} ratio as a function of
$L_{\rm FIR}$, for a sample of nearby 
normal and starburst galaxies 
(Malhotra et al. 2001), local ULIRGs (Farrah et al.\, 2013, Diaz-Santos
et al.\, 2013), high-$z$ star-forming
and AGN-powered sources  (Stacey et al.\,2010), Hailey-Dunsheath
et al.\, 2010), high-$z$ lenses from Ivison et al (2010), Valtchanov et
al. (2011), George et al. (2013) and our
sample. Local
star forming and normal galaxies and high-redshift star forming
galaxies display ratios of {\it R}$\sim$0.001$-$0.01 while {\it R} is
in the range {\bf (1.3$-$10)}$\times$10$^{-4}$ for local ULIRGs. 
The ratio varies between {\it R}$\approx$ 3$\times$10$^{-4}$$-$
0.7$\times$10$^{-2}$ for our sample (U)LIRGs. 
Turning to the two AGN sources in our sample with low  $L_{\rm [CII]}/L_{\rm FIR}$ we note that it is plausible that a
sizeable fraction of the $L_{\rm FIR}$ originates in the AGN component
hence lowering the overall $L_{[\rm CII]}/L_{\rm FIR}$ ratio.

Our finding reveals, for the first time, that intermediate redshift
(U)LIRGs with no AGN have 
$L_{\rm [CII]}/L_{\rm FIR}$ ratios similar to those of high-redshift
star-forming galaxies and local normal galaxies. The implications
of this result  for the nature and evolution of ULIRGs are discussed in Section 4.

\subsection{The [CII]$/$CO(1$-$0) ratio and PDR properties}

Early [CII] surveys of extragalactic sources and mapping of the Galaxy
(e.g. Heiles et al. 1994, Cubick et al. 2008) have established that a
large fraction of [CII] emission  in galaxies originates in PDRs on the outer
layers of molecular clouds exposed to intense far-UV
radiation. [CII] also acts as a coolant of the low density warm
ionized medium. Recent studies (e.g. Rigopoulou et al. 2013) 
found that up to 25\% of the [CII] emission in the star
forming galaxy IC342 originates in the diffuse ionized gas. 
We will thus assume that the majority of [CII]
emission in our sample (U)LIRGs originates in PDRs.
Kaufmann et al. (1999) presented PDR models in which the line emission
from the clouds is determined by the density of the gas $\eta$ and
the incident flux ${\it G_{0}}$ (expressed in units of
the Habing Field, 1.6$\times$ 10$^{-3}$ erg cm$^{-2}$ s$^{-1}$). The
models assume that the [CII] line and the FIR continuum are optically
thin  optically thick and self-absorption case is not likely (see e.g. Luhman et al. 2003)
but, the low-J CO transitions are optically thick.
So when comparing models to observables one needs to subtract off the
fraction of the [CII] line arising in the ionized medium 
($\sim$ 25\%, e.g. Rigopoulou et al.\, 2013), and multiply the CO line intensity by a factor of
two for all objects.


Using the CO measurements reported in  
Magdis et al. (2014) we examine the $L_{\rm [CII]}/L_{\rm CO(1-0)}$
ratio for
our intermediate redshift ULIRGs. For sources without CO(1$-$0) measurements we use 
the conversion factors for submm galaxies CO(2-1)$/$CO(1-0) $\sim$0.84 $\pm$ 0.13 and 
CO(3-2)$/$CO(1-0) $\sim$ 0.52 $\pm$ 0.09. (e.g.~Bothwell et al.\, 2013).
The $L_{\rm [CII]}/L_{\rm CO(1-0)}$ ratio for our sample is 3300
$\pm$ 420. This value excludes the two sources that contain
confirmed AGN. 
For comparison, local ULIRGs have a mean
$L_{\rm [CII]}/L_{\rm CO(1-0)}$ ratio of 1500$\pm$260,  high-z
star-forming galaxies 4050$\pm$410, normal local galaxies 1800$\pm$270, while
starburst nuclei and local Galactic star-forming regions have a ratio
of 4100$\pm$320 (Stacey et al.\, 1991). 
We find that the mean
$L_{\rm [CII]}/L_{\rm CO(1-0)}$ value
for our sample is a factor of two larger
than that of local ULIRGs and closer to the value found for
local and high-$z$ star-forming galaxies. 
Although 
local normal galaxies and local ULIRGs have similar
$L_{\rm [CII]}/L_{\rm CO(1-0)}$ ratios, their $L_{\rm
 [CII]}/L_{\rm FIR}$ ratios
are a factor of 10 different. 
Local normal galaxies
have higher  
$L_{\rm [CII]}/L_{\rm FIR}$
values than those seen in local ULIRGs. 
Likewise, the
$L_{\rm [CII]}/L_{\rm CO(1-0)}$ and $L_{\rm [CII]}/L_{\rm FIR}$ ratios of
local and intermediate redshift ULIRGs are different by a factor of 2
and 10, 
reinforcing our earlier findings that the properties of the
ISM of ULIRGs changes dramatically between z=0 and z=0.5.

To further investigate these trends, in Figure 3 we plot the 
$L_{\rm [CII]}/L_{\rm FIR}$ and $L_{\rm CO(1-0)}/L_{\rm FIR}$ ratios for our
sample and compare them to those of galactic star-forming regions, 
local ULIRGs, local normal galaxies, high and low-$z$ star-forming
galaxies and  recent {\it Herschel} measurements
of high-$z$ lensed star-forming galaxies from Ivison et
al (2010), Valtchanov et al. (2011), and George et al. (2013). 
For reference,  we also show PDR model calculations\footnote{based on
PDRToolbox (PDRT) web site: http:$//$dustem.astro.umd.edu$/$)}
for gas density $\eta$ and FUV strength $G_{0}$ (adapted from
Hailey-Dunsheath et al 2010) applicable to sources without a dominant
AGN component.


The mean 
$L_{\rm [CII]}/L_{\rm FIR}$ and $L_{\rm CO(1-0)}/L_{\rm FIR}$ 
ratios for intermediate redshift ULIRGs are a factor of 10 and 7
times higher than those of local ULIRGs.
Since the emergent
$L_{\rm FIR}$ is proportional to the incident radiation flux
(parameterised by $G_{0}$) the smaller $L_{\rm [CII]}/L_{\rm FIR}$
line ratios observed in local ULIRGs indicate that [CII] must be produced in dense PDRs
illuminated by a strong radiation field (e.g. Farrah et al. 2013).
In a recent study of [CII] line emission from a
sample of local ULIRGs Diaz-Santos et al. (2013) argue in favor of 
smaller$/$compact star-forming regions.

Clearly, none of the above scenarios are applicable to the
intermediate redshift (U)LIRGs. 
The data points for our sample fall in the range 10$^{3}<G_{0}<10^{2}$
an area which overlaps with the low end of the distribution of the
{\it G$_{0}$} values for starbursts and the high end of $G_{0}$ values for normal galaxies. Hence,
we conclude that the bulk of the molecular gas in intermediate
redshift (U)LIRGs is exposed to a softer radiation field than that of
local ULIRGs, more akin to those found in nearby starburst nuclei.

The $L_{\rm CO(1-0)}/L_{\rm FIR}$ of intermediate redshift (U)LIRGs is
$\sim$7 times
higher than that of local ULIRGs. It is similar to the upper end
of the values seen in local starburst nuclei and local normal
galaxies. However, as we discussed earlier, intermediate redshift
ULIRGs display an $L_{\rm [CII]}/L_{\rm CO(1-0)}$ ratio of 3000 which is 
higher than the value found in local normal galaxies but lower that
that of starburst nuclei (4100). A lower
$L_{\rm [CII]}/L_{\rm CO(1-0)}$ 
(for the same $L_{\rm CO(1-0)}/L_{\rm FIR}$ value) 
ratio would imply that most of the CO emission originates in molecular
clouds residing in less active star-forming regions than those
responsible for producing the fine-structure emission lines.

\section{The Nature and Evolution of Luminous Infrared Galaxies}

The star formation rate density (SFRD) in the Universe increases dramatically
from the present day to $z$$\sim$1 at which point it 
becomes flat out to  $z$$\sim$3$-$4
(e.g. Bouwens et al. 2009, Magnelli et al. 2013). 
Luminous Infrared Galaxies are a dominant component to the co-moving
SFRD between $z$$\sim$0.5$-$1 so their properties may give clues to the drivers behind the dramatic rise at $z$$<$1.

Although local ULIRGs have been used as templates for $z$$\sim$2 luminous infrared
galaxies, increasing
evidence exists (e.g. Papovich et al. 2007, Farrah et al. 2008, Swinbank et
al. 2010) that they are distinctly different to their high-$z$
counterparts. 

It is now well established that the intense
star-forming activity of local ULIRGs is the result of merging
(e.g. Sanders \& Mirabel 1999). Morphology indications in 
high-resolution {\it HST} images however, reveal that, at $z$$\sim$2 at least 50\% of the
Herschel-selected starbursts are not driven by galaxy mergers
(e.g. Kartaltepe et al. 2011). In section 3 we presented new evidence that
the properties of the star-forming regions of our intermediate
redshift (U)LIRGs are distinctly different to those of local
ULIRGs. Instead, they bear a closer resemblance to those found in high redshift star-forming galaxies.
In particular,
we found evidence that the star-forming regions of our $z$$\sim$0.5 (U)LIRGs 
are illuminated by moderately intense FUV radiation, with $G_{0}$ in
the range 10$^{2}$$-$10$^{2.5}$ and ${\it G_{0}}/ \eta \sim$0.1$-$1
cm$^{3}$ (when the appropriate corrections
of 0.7$\times$$L_{\rm[CII]}$ and 2$\times$$L_{\rm CO}$ are applied). 
Similar  $G_{0}/ \eta$ values have been found for
local normal and starburst galaxies (e.g. Wolfire et al. 1990).
In contrast, the low $L_{[\rm CII]}/L_{\rm FIR}$ values seen in local ULIRGs
are the result of the presence of  intense radiation fields ${\it G_{0}}$ with high column
densities ${\it G_{0}}/ \eta \sim$ 0.02$-$0.03 cm$^{3}$ (Farrah et
al. 2013) and possibly
compact sizes (e.g. Diaz-Santos et al. 2013).
The high $L_{\rm [CII]}/L_{\rm FIR}$ values seen
in intermediate redshift ULIRGs together with the modest radiation
fields speak against the existence of such dense compact PDRs, hence the
star-forming regions of intermediate redshift (U)LIRGs
must be extended. 

Our survey has revealed  a strong evolution in 
the properties of  ULIRGs at 0.2$<$z$<$0.8 and confirms the use
of the [CII] line  and the $L_{\rm [CII]}/L_{\rm IR}$  and
 $L_{\rm C[II]}/L_{\rm CO(1-0)}$ ratios,
in probing the properties of the star forming regions 
in galaxies at low, intermediate and high redshifts. 
Even at modest redshifts (z$\sim$0.5) the
nature of the ULIRG population changes significantly from exclusively
compact merger-driven to a more varied population.

\acknowledgments{
We thank the anonymous referee for his$/$her insightful comments.
DR and GEM acknowledge support from grant 
ST$/$K00106X$/$1 and the John Fell Oxford University
Press (OUP) Research Fund (GEM).
This paper is based on data from {\it Herschel's} SPIRE-FTS. 
SPIRE has been developed by a consortium of institutes led by
Cardiff Univ. (UK) and including: Univ. Lethbridge (Canada);
NAOC (China); CEA, LAM (France); IFSI, Univ. Padua (Italy);
IAC (Spain); Stockholm Observatory (Sweden); Imperial College
London, RAL, UCL-MSSL, UKATC, Univ. Sussex (UK); and
Caltech, JPL, NHSC, Univ. Colorado (USA). This development
has been supported by national funding agencies: CSA (Canada);
NAOC (China); CEA, CNES, CNRS (France); ASI (Italy); MCINN
(Spain); SNSB (Sweden); STFC, UKSA (UK); and NASA (USA).

{\it Facilities:} \facility{Herschel}, \facility{IRAM:30m}, \facility{APEX}.





\begin{figure*}
\centering
\includegraphics[scale=0.5,angle=-90]{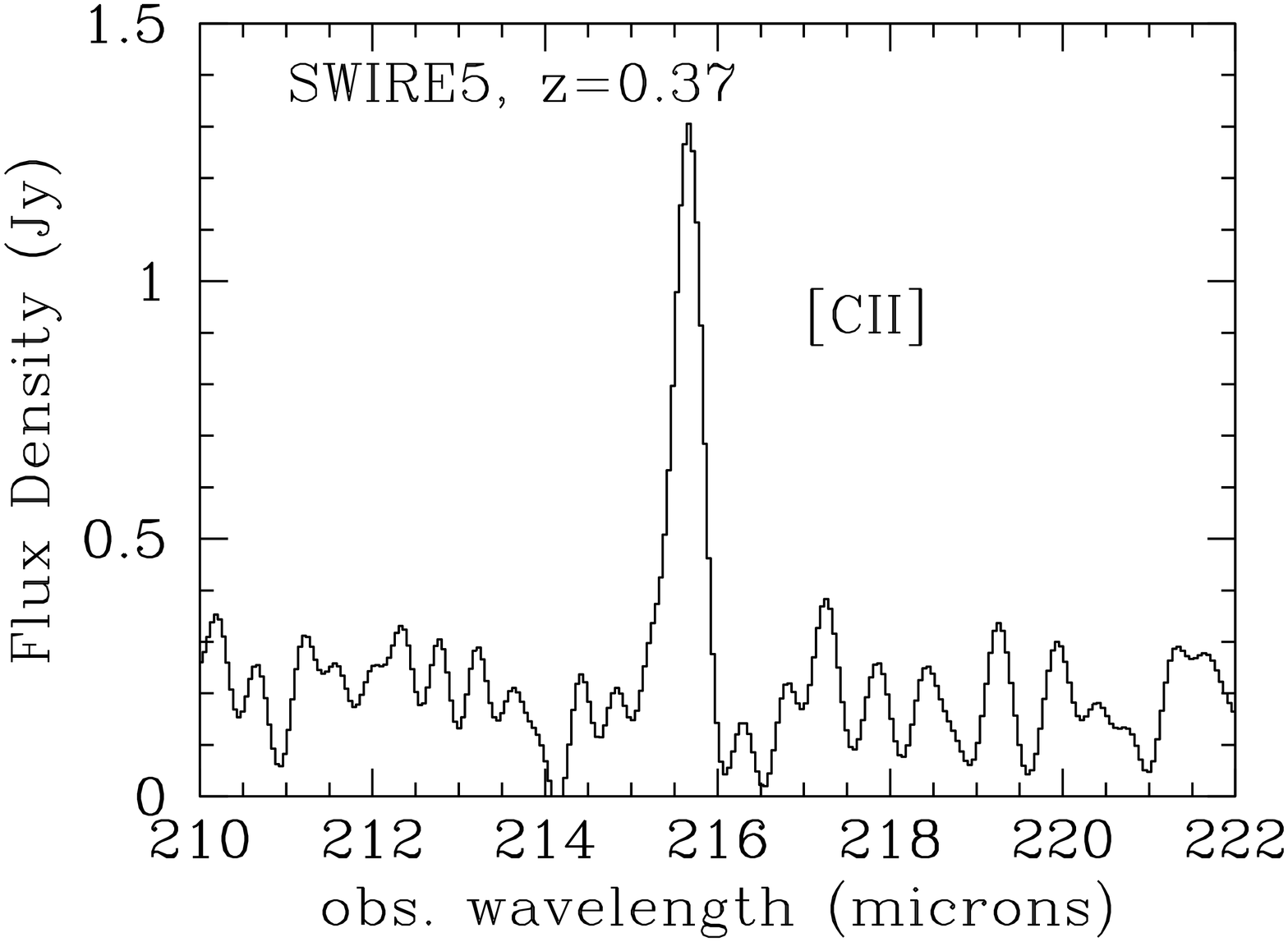}
\caption{Continuum-subtracted region of the FTS spectrum around the
[CII] line in the observed frame for one of our sample ULIRGs (SWIRE5) at z=0.37. }
\end{figure*}

\begin{figure*}
\centering
\includegraphics[scale=0.5]{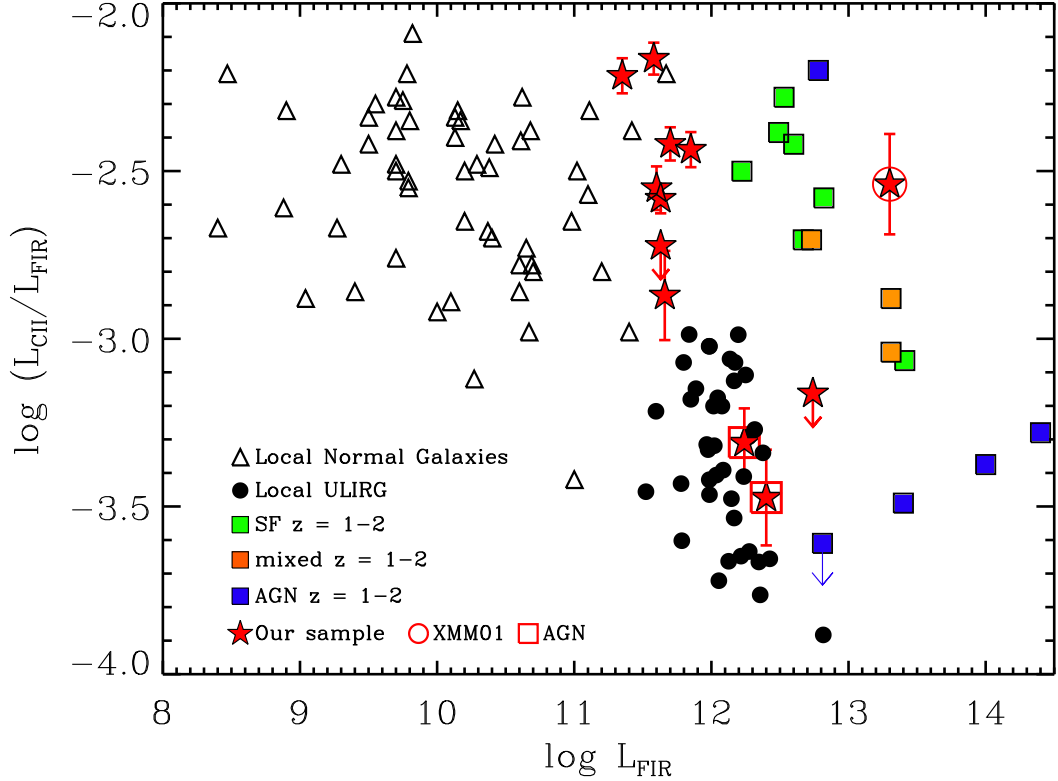}
\caption{Log($L_{[\rm CII]}/L_{\rm FIR}$) as a function of log($L_{\rm FIR}$)
  for local normal and starburst galaxies (open triangles) from
  Malhotra et al. (2001), local
  ULIRGs (black filled circles) from Farrah et al. (2013) and
  Diaz-Santos et al. (2013), high-redshift (1 $<$ z $<$ 2)
  star-forming galaxies (green squares), mixed systems (orange filled
  squares) and AGN (blue filled squares) from Stacey et al. (2010). 
Red stars denote the 12 intermediate redshift (U)LIRGs presented in this
  work (circled star is XMM01, squared stars denote AGN).  Note that  $L_{\rm FIR(42.5-122.5\,\mu m)}$ = 0.63 $\times$ $L_{\rm IR(8-1000 \mu m)}$.}
\end{figure*}

\begin{figure*}
\centering
\includegraphics[scale=0.5]{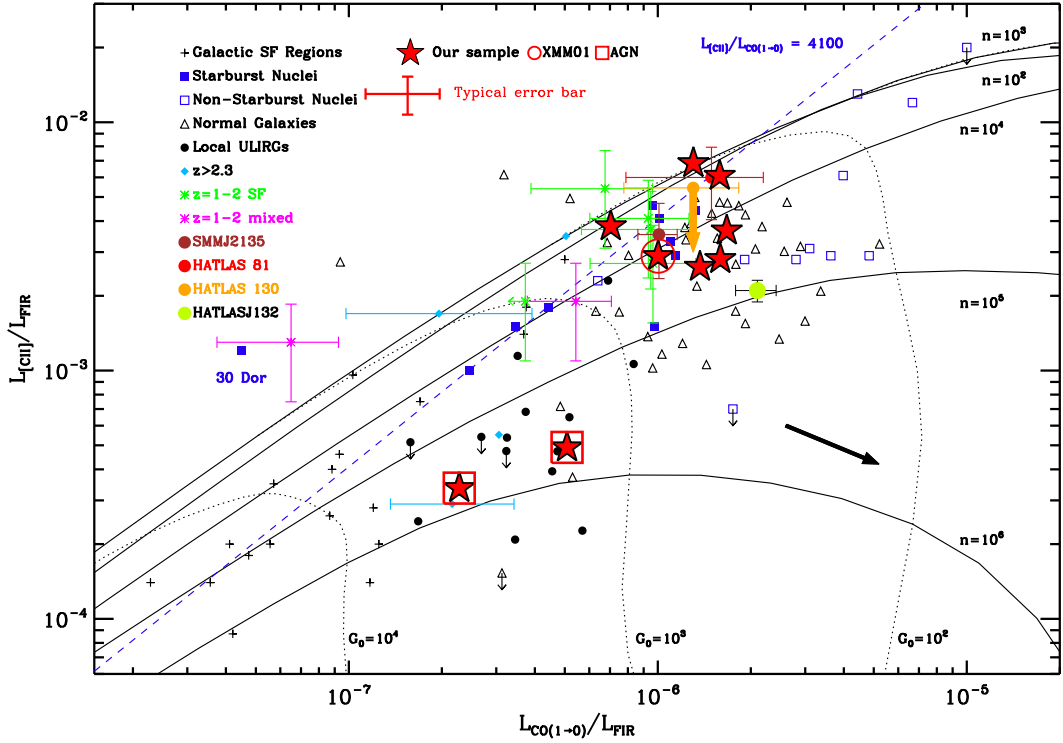}
\caption{$L_ {\rm [CII]}/L_{\rm FIR}$ as a function of $L_{\rm CO(1-0)}/L_{\rm FIR}$ for Galactic
  star-forming regions (crosses), local starburst nuclei (filled
  squares), local non-starburst nuclei (open squares), local normal
  galaxies (triangles), local ULIRGs (circles), redshift 1 to 2
  sources (asterisks with error bars), high z sources (cyan
  diamonds), adapted from Hailey-Dunsheath et al. (2010).
   Herschel lenses are
  from: SMMJ2135 (Ivison et al. 2010), HATLAS 81 \& HATLAS 130
  (Valtchanov et al. 2011), HATLASJ132 (George et al. 2013). 
 Red asterisks denote the present sample (circled star is XMM01,
 squared stars denote AGN).
The typical error bar shown refers to our sample only.
 Overplotted are the PDR model values for
  $\eta$ and $G_{0}$ from Kaufmann et al. (1999). The plot is
  based on observed values and is intended as a first order diagnostic
  tool, hence does not include any correction. The black arrow
  indicates the direction that all data points in the plot will shift
  when corrections are applied (see text).} 
\end{figure*}

\end{document}